\documentclass[12pt]{iopart}

\usepackage[dvips]{graphicx}
\usepackage{mathrsfs}  

\newcommand{\ie}{\emph{i.e.} }
\newcommand{\eg}{\emph{e.g.} }

\DeclareRobustCommand{\Eqref}[1]{Eq.~(\ref{#1})}
\DeclareRobustCommand{\Figref}[1]{Fig.~(\ref{#1})}

\begin{document}

\title{The Peierls argument for higher dimensional Ising models}

\author{Claudio Bonati}
\address{Dipartimento di Fisica, Universit\`a di Pisa and INFN, Sezione di Pisa, 
Largo Pontecorvo 3, I-56127 Pisa, Italy.}
\ead{bonati@df.unipi.it}

\begin{abstract}
The Peierls argument is a mathematically rigorous and intuitive method to show the 
presence of a non-vanishing spontaneous magnetization in some lattice models. 
This argument is typically explained for the $D=2$ Ising model in a way which cannot be easily generalized
to higher dimension. The aim of this paper is to present an elementary discussion of the Peierls 
argument for the general $D$-dimensional Ising model.
\end{abstract}


\maketitle

\section{Spontaneous symmetry breaking and the Ising model}

Spontaneous symmetry breaking is a unifying theme of modern theoretical physics, with applications
ranging from statistical physics \cite{HuangBook} to condensed matter \cite{AndersonBook, CLBook}
and particle physics \cite{Weinberg2Book}. Because of its fundamental relevance in our understanding
of nature, it is important to study simple systems for  
which the presence of spontaneous symmetry breaking can be rigorously established.
Probably the simplest system which displays spontaneous simmetry breaking is the Ising model in 
statistical physics, named after E.~Ising, who first solved the unidimensional version of the problem \cite{Ising25}.

This model is defined on a graph, \ie on a set of points (called sites in the following) 
equipped with the notion of nearest neighbourhood. 
A variable $s_i$, which takes value in the set $\{+1,-1\}$, is associated to each site $i$ of the graph. 
We will call ``configuration'' a given assignment of the variables $\{s_i\}$ to the graph sites
and the energy of a configuration is defined by
\begin{equation}\label{E}
E=-J\sum_{(i,j)} s_is_j-h\sum_i s_i\ ,
\end{equation}
where $J, h$ are constants and $\sum_{(\cdot,\cdot )}$ denotes the sum on the first neighbour 
sites. For the sake of the simplicity in the following we will only consider the model defined on an 
hypercubic lattice in $D$ dimensions, which is the most studied case. 

For $J>0$ this model describes an uniaxial ferromagnet in an external magnetic field of intensity $h$. 
Configurations in which most of the nearest neighbour sites are oriented in the same direction 
are favoured, since they correspond to lower values of the energy. 
Moreover configurations with the value $\mathrm{sign}(h)$ in most sites 
are favoured by the interaction with the external magnetic field. 

The model with $h=0$ is particularly interesting since the energy is invariant under the transformation
\begin{equation}\label{sym}
s_i\to -s_i \quad \forall i\ . 
\end{equation}
By applying this transformation two times we come back to the original configuration, so the symmetry 
group of the model for $h=0$ is $Z_2$. In the following of this paper we will be interested just in this 
$h=0$ case.

We now introduce the average magnetization per site of a configuration, defined by
\begin{equation*}
m=\frac{1}{N}\sum_i s_i=\frac{N_+-N_-}{N}\ , 
\end{equation*}
where $N$ is the total number of sites and $N_{\pm}$ is the number of sites with  $s_i=\pm 1$. 
It is clear from this definition that $m$ is odd under the transformation in \Eqref{sym}, \ie $m$ goes to $-m$. 
Since the energy of two configurations related by the symmetry in \Eqref{sym} is the same, one could think 
that the statistical average of $m$, denoted by $\langle m\rangle$, identically vanishes.

This is true for finite systems, however, in the thermodynamical limit, 
if $D\ge 2$ the magnetization $\langle m \rangle_{\infty}$ vanishes only in the high temperature paramagnetic phase. 
In the low temperature ferromagnetic phase the value of $\langle m\rangle_{\infty}$ is not 
well defined and depends on how the thermodynamical limit is performed.
In this case the symmetry in \Eqref{sym} is said to be spontaneously broken. 

The breaking of a symmetry can be thought as a form of thermodynamical instability: the particular 
value acquired by $\langle m\rangle_{\infty}$ in the ferromagnetic phase is determined by small 
perturbations. A conventional way to uniquely define $\langle m\rangle_{\infty}$ in the broken phase (where 
it is called spontaneous magnetization) is to use an infinitesimal magnetic field:
\begin{equation}\label{spont_mag}
\langle m\rangle_{\infty}=\lim_{h\to 0^+}\lim_{N\to\infty} \langle m\rangle \ ,
\end{equation}
where it is crucial to perform the thermodynamical limit before switching off the magnetic field. 
The instability manifests itself in that using $h\to 0^-$ in \Eqref{spont_mag} would change the sign of 
$\langle m\rangle_{\infty}$.

A different approach to expose the instability is the use of appropriate boundary conditions: we can for 
example impose in all the sites $i_b$ on the lattice boundary the condition $s_{i_b}=+1$. In the 
paramagnetic phase the effect of the boundary conditions does not survive the thermodynamical limit, 
while in the ferromagnetic phase their effect is analogous to that of the infinitesimal magnetic 
field in \Eqref{spont_mag}. 

On a finite lattice the mean value of the magnetization $m$ can be written in the form
\begin{equation}\label{mmean}
\langle m\rangle=\frac{\langle N_+\rangle -\langle N_-\rangle }{N}=1-2 \frac{\langle N_-\rangle}{N}\ ,
\end{equation}
where we used the fact $N_++N_-=N$, and in order to show that $\langle m\rangle_{\infty}>0$ 
it is sufficient to show that for every $N$ we have $\langle N_-\rangle/N <1/2-\epsilon$ 
(with $\epsilon>0$ and $N$-independent). 
The Peierls argument is a simple geometrical construction that can be used to prove this bound. 
It was introduced for the first time in \cite{Peierls36} and some errors in the estimates used were later 
corrected in \cite{Griffiths64}. The original formulation referred to the two dimensional Ising model, whose 
solution \cite{Onsager44} was still not known, but the idea of the argument can be adapted also 
to the general $D$ dimensional problem with $D>2$, which is still an active field of research 
(see \eg \cite{boot1, boot2, boot3}). 

The outcome of the Peierls argument for the model in $D$ dimensions is an estimate of the form 
\begin{equation}\label{generalbound}
\langle N_-\rangle \le N f_D(x)\, 
\end{equation}
where $x$ is defined by
\begin{equation}\label{x}
x=9e^{-4J\beta} \qquad \beta=1/(kT)
\end{equation}
and $f_D(x)$ is a continuous function of $x$ (independent of $N$) such that $\lim_{x\to 0} f_D(x)=0$. In particular for 
small enough $T$ we have the bound $\langle N_-\rangle /N<1/2-\epsilon$ ($\epsilon>0$), which ensures 
that $\langle m\rangle_{\infty}\ge2\epsilon$ and the $Z_2$ symmetry is spontaneously broken.

The original $D=2$ argument is described in most books on statistical mechanics (like \eg 
\cite{HuangBook}, \S 14.3), however the construction is presented in such a way that the 
generalization to higher dimension is not immediate: the use of ordered paths in $D=2$ simplifies 
the proof of some of the estimates but cannot be easily generalized to the higher dimensional setting. Moreover the 
combinatorics needed for $D>2$ appears at first sight to be much more involved than the one required 
in the two dimensional case. 
On the other hand the higher dimensional problem is discussed in specialized books  (like 
\cite{RuelleBook, SinaiBook, SimonBook}), but the topic is approached from a different and more abstract point of view, 
out of reach for most of the students of a first course in statistical mechanics. The consequence could be the (erroneous!) 
feeling that the Peierls argument can be conveniently applied only in the case in which, strictly speaking, it is no more
necessary, being the two-dimensional Ising model exactly solvable.

The purpose of this paper is to fill this gap by presenting an elementary discussion of
the Peierls argument in its more general $D$-dimensional version.
First of all we present the Peierls argument in $D=2$ using a construction scheme of the domains different from the 
one typically adopted, which can be easily generalized to the higher dimensional case, then we go on to show how to solve the
additional problems that arise in the higher dimensional environment. With this aim we discuss in some detail the 
$D=3$ problem, where one still has a good geometrical intuition, and finally analyze the general $D$ dimensional 
case, which is at this point an almost trivial extension of the three dimensional one.

\section{The $D=2$ problem}

\begin{figure}[t]
\begin{center}
\includegraphics[width=0.4 \textwidth, clip]{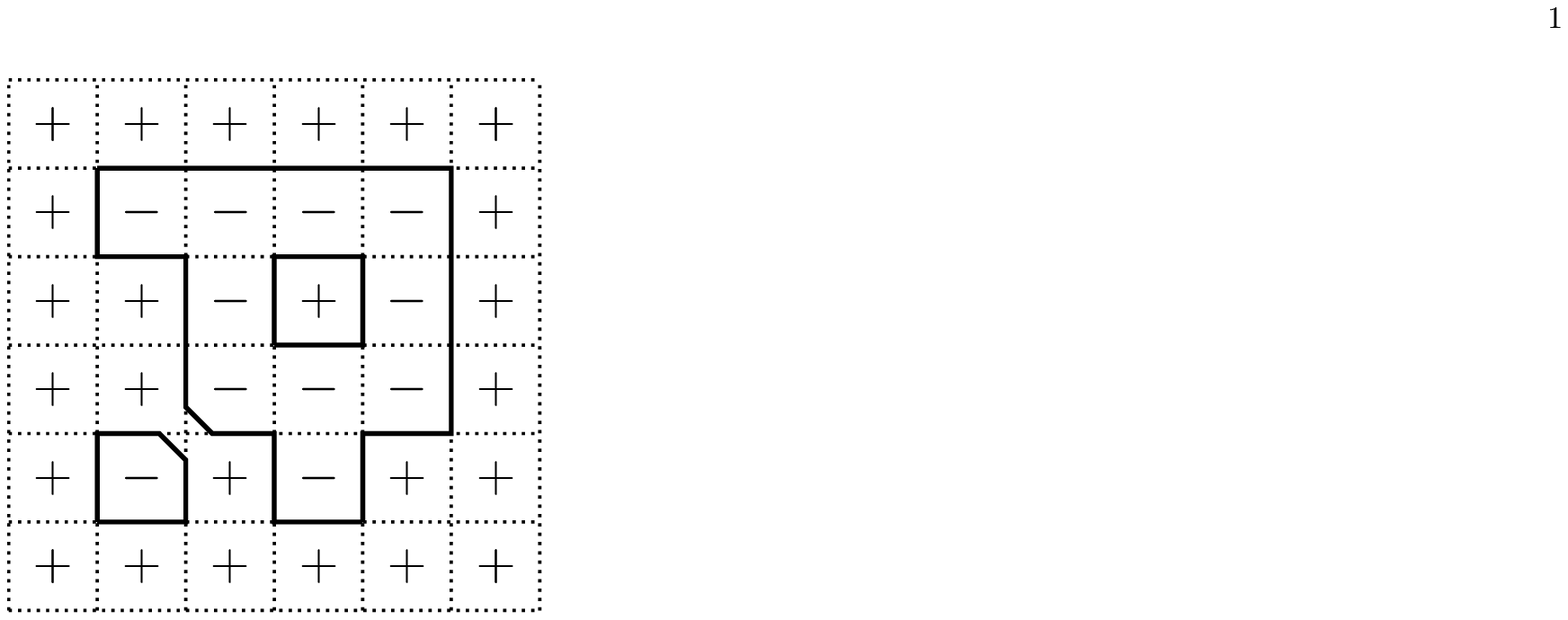}
\hspace{1cm}
\includegraphics[width=0.4 \textwidth, clip]{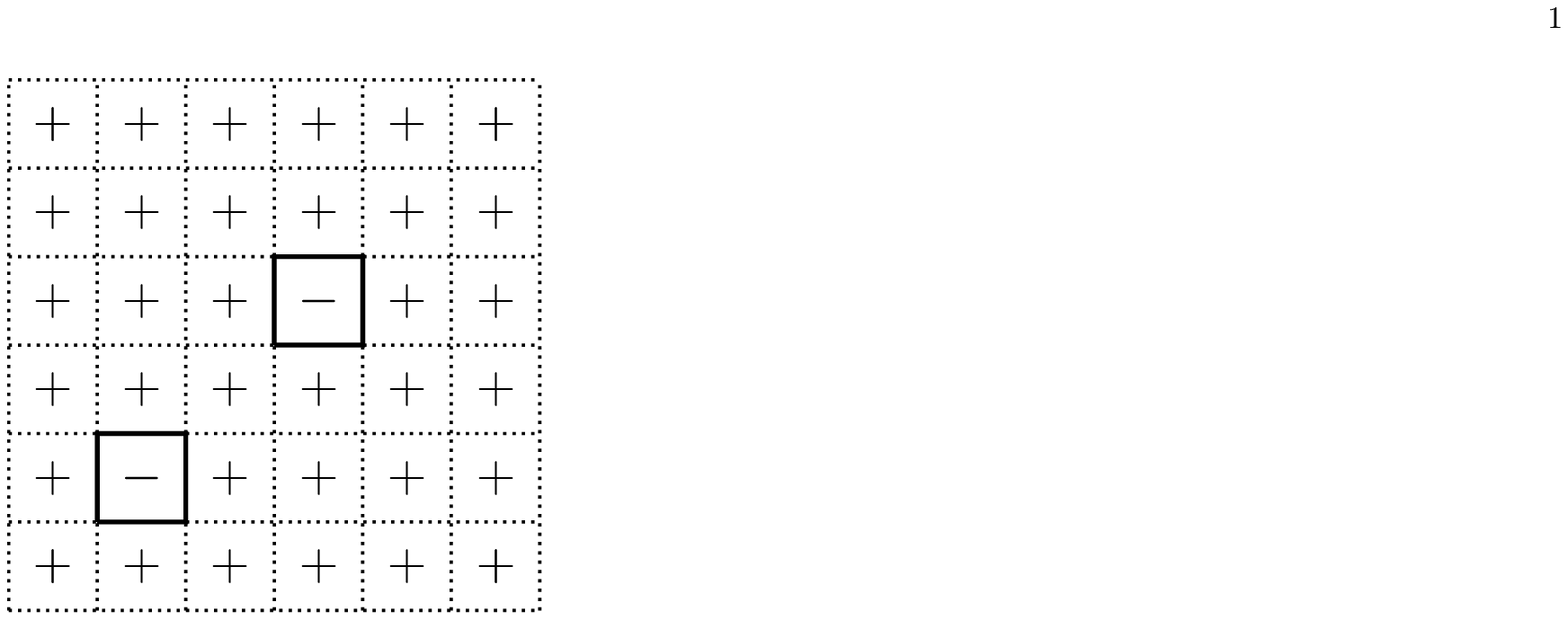}
\end{center}
\caption{(\emph{left}) An example of construction of the Peierls contours.
(\emph{right}) An example of the transformation $c\to\tilde{c}$ (see text), in which
$\gamma_L^i$ is the longest contour of the configuration shown in the left panel.}\label{2dconstr_transf}
\end{figure}

We consider a two dimensional square lattice of size $\sqrt{N}\times \sqrt{N}$ (with 
lattice spacing $a=1$) and fix $s_{i_b}=+1$ on the boundary sites $i_b$. The Peierls 
contours are introduced by the following procedure:
\begin{enumerate}
\item draw a unit square on each site $i$ with $s_i=-1$
\item cancel the edges that appear twice (\ie that separate two neighbour sites $i,j$ with $s_i=s_j=-1$)
\item in the case in which four edges meet at the same point, chop off the corner 
of the squares in order to remove ambiguities.
\end{enumerate}
An example of the application of this procedure is shown in \Figref{2dconstr_transf} and 
it is immediate to show that the following facts are true:
\begin{itemize}
\item every contour is a closed non-intersecting curve
\item every site $i$ with $s_i=-1$ is inside at least one contour
\item the set of the admissible contours is in a one-to-one
      correspondence with the set of the configurations.
\end{itemize}
The one-to-one relation in the last property depends on the fixed $s_{i_b}=+1$ 
boundary conditions: the presence of a contour signals a change of sign of the 
site variable and the $+1$ assignment on the boundaries 
uniquely fix the signs.

In a finite lattice the number of contours of given length $L$ is finite, let us 
denote this number by $\#(L)$. The generic Peierls contour can then be 
denoted by $\gamma_{L}^{i}$, where $L$ is the length of the contour and 
$1\le i\le \#(L)$. If $A(\gamma)$ is the area of the contour $\gamma$ 
(that is the number of sites inside the contour) we have the following upper 
bound for the number $N_-$ of sites with $s_i=-1$ present in a configuration:
\begin{equation}\label{2d_0}
N_-\le \sum_{L\ge 4, \mathrm{even}}\sum_{i=1}^{\#(L)} A(\gamma_{L}^i)X(\gamma_L^i)
\end{equation}
where $X(\gamma)=1$ if $\gamma$ occurs in the given configuration and $X(\gamma)=0$ 
otherwise. The lower bound of the sum is $4$ since this is the minimal length possible
for a closed contour, moreover the length of a closed contour has to be even, hence 
the sum extends on even numbers only. 
\Eqref{2d_0} is a simple consequence of the fact 
that every site $i$ with $s_i=-1$ is inside at least one contour: if we sum the
area of all the contours that occur in a configuration we get an upper bound for $N_-$
in that configuration.

\begin{figure}[t]
\begin{center}
\includegraphics[width=0.4 \textwidth, clip]{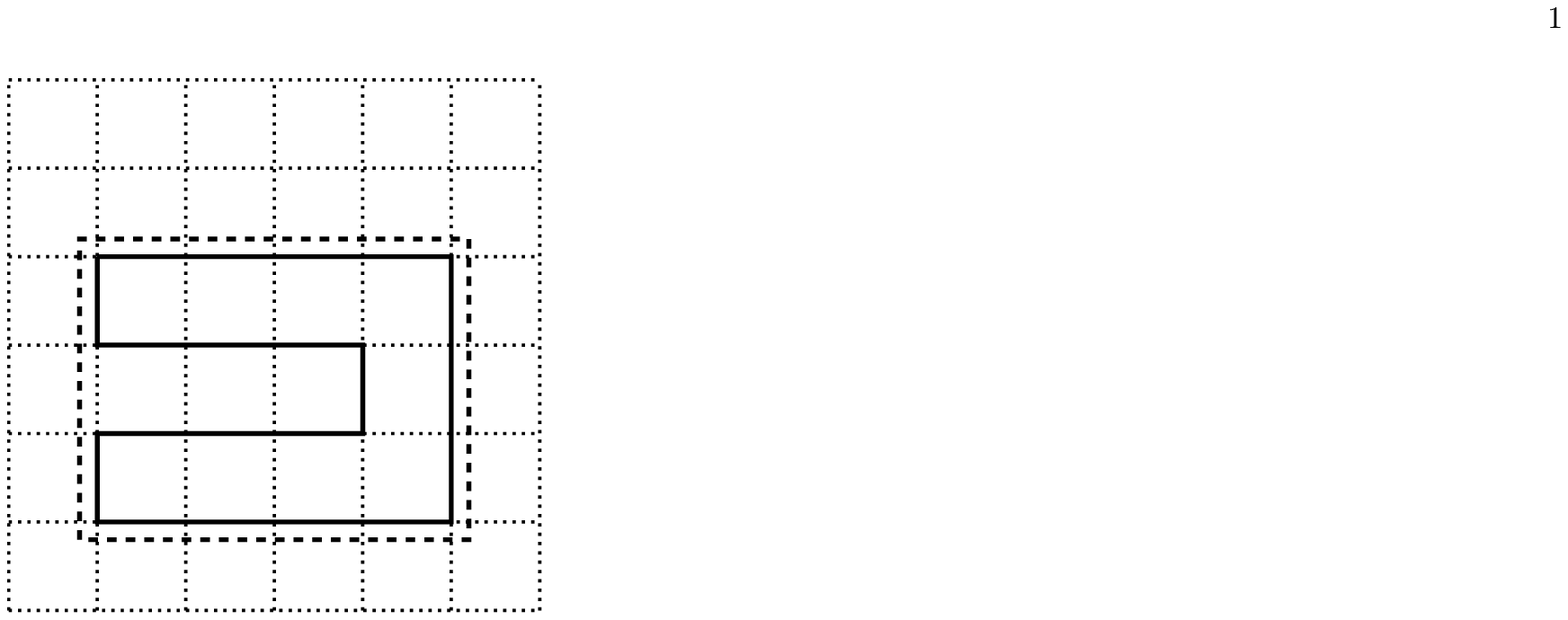}
\end{center}
\caption{The first step of the proof of the bound for $A(\gamma_L^i)$ (the edges of the 
rectangle are slightly shifted to improve readability).}\label{2d_AL_proof_fig}
\end{figure}

The next step is to show that $A(\gamma_L^i)$ has an upper bound of the form $A(L)$, 
\ie that depend only on the length $L$ of the contour. This upper bound can be 
obtained in the following way: draw the smallest rectangle $\mathcal{R}$ (with 
sides parallel to the boundaries of the lattice) that contain $\gamma_L^i$ (see 
\Figref{2d_AL_proof_fig} for an example). The perimeter of $\mathcal{R}$ is 
not larger than $L$: if we draw inside $\mathcal{R}$ a line parallel to one of the axes, 
this line intersects $\mathcal{R}$ in two edges, but this line also has 
to intersect $\gamma_L^i$ in at least two edges, otherwise $\gamma_L^i$ would 
be separated in two disjoint contours or $\mathcal{R}$ would not be the smallest 
rectangle containing $\gamma_L^i$. If we denote the length of the sides of 
$\mathcal{R}$ by $x_1$ and $x_2$, we thus have $2(x_1+x_2)\le L$ 
and $A(\gamma_L^i)\le x_1x_2$. As a consequence 
\begin{equation}\label{2d_AL_part}
A(\gamma_L^i)\le \max_{2(x_1+x_2)\le L} x_1x_2\ . 
\end{equation}
It is simple to show that the maximum in the previous equations is reached when 
$x_1=x_2=L/4$ (\ie of all the rectangles of fixed perimeter the square is the one with 
the largest area) and we conclude that 
\begin{equation}\label{2d_AL}
A(\gamma_L^i)\le A(L)\equiv \frac{L^2}{16}\ .
\end{equation}
We can now use \Eqref{2d_AL} to modify \Eqref{2d_0} as follows:
\begin{equation*}
N_-\le \sum_{L\ge 4, \mathrm{even}} A(L) \sum_{i=1}^{\#(L)} X(\gamma_L^i)\ .
\end{equation*}

In the following we will not be interested on the value of $N_-$ for a single 
configuration, but on the mean value $\langle N_-\rangle$. From the previous 
equation we get
\begin{equation}\label{2d_1}
\langle N_-\rangle \le \sum_{L\ge 4, \mathrm{even}}A(L) \sum_{i=1}^{\#(L)} 
\langle X(\gamma_L^i)\rangle 
\end{equation}
and our next task will be to show that $\langle X(\gamma_L^i)\rangle \le X(L)$, 
where the function $X(L)$ depends only on the length $L$ of the path $\gamma_L^i$. 

The thermal average $\langle X(\gamma_L^i)\rangle$ is defined by
\begin{equation*}
\langle X(\gamma_L^i)\rangle = \frac{\sum_{c\in\mathscr{B}} 
X(\gamma_L^i)e^{-\beta E(c)}}{\sum_{c\in\mathscr{B}} e^{-\beta E(c)}}\ ,
\end{equation*}
where $\beta=1/(kT)$ and the sum is extended over the set $\mathscr{B}$ of configurations that satisfy the $s_{i_b}=+1$ 
boundary conditions. From the definition of $X(\gamma_L^i)$ 
we immediately see that its mean value can be rewritten in the form
\begin{equation*}
\langle X(\gamma_L^i)\rangle = \frac{\sum_{c\in\mathscr{C}}\, e^{-\beta E(c)}}{
\sum_{c\in\mathscr{B}} e^{-\beta E(c)}}\ ,
\end{equation*}
where $\mathscr{C}$ is the set of all the configurations in $\mathscr{B}$ which contain the Peierls 
contour $\gamma_L^i$. Let $c$ be a configuration in $\mathscr{C}$ and 
define the configuration $\bar{c}$ as the one obtained from $c$ by changing the sign of all 
the variables associated to sites inside $\gamma_L^i$ (an example of such a transformation is given in 
\Figref{2dconstr_transf}). It is clear that $\bar{c}\in\mathscr{B}$, since the boundary 
values of the configuration are unchanged, so the set of all the possible 
$\bar{c}$ is a subset of the collection $\mathscr{B}$: 
\begin{equation*}
\bar{\mathscr{C}}=\{\bar{c}|c\in\mathscr{C}\}\subset \mathscr{B} 
\end{equation*}
and thus
\begin{equation*}
\sum_{\bar{c}\in\bar{\mathscr{C}}} e^{-\beta E(\bar{c})}\le 
\sum_{c\in\mathscr{B}} e^{-\beta E(c)}\ .
\end{equation*}
From this relation see that
\begin{equation*}
\langle X(\gamma_L^i)\rangle \le \frac{\sum_{c\in\mathscr{C}}\, e^{-\beta E(c)}}{
\sum_{\bar{c}\in\bar{\mathscr{C}}} e^{-\beta E(\bar{c})}}
\end{equation*}
and, noting that if $c\in\mathscr{C}$ we have
\begin{equation}\label{2d_energy}
E(c)=E(\bar{c})+2JL\ ,
\end{equation}
we obtain the upper bound
\begin{equation}\label{2d_X}
\langle X(\gamma_L^i)\rangle \le X(L)\equiv e^{-2J\beta L}\ .
\end{equation}
By using this bound in \Eqref{2d_1} we get
\begin{equation}\label{2d_2}
\langle N_-\rangle \le \sum_{L\ge 4, \mathrm{even}} A(L) \sum_{i=1}^{\#(L)} X(L)
= \sum_{L\ge 4, \mathrm{even}} A(L) \#(L) X(L) 
\end{equation}
and to finish we need an upper bound for $\#(L)$, \ie for the number of 
closed paths of length $L$. This can be obtained by enumerating the possible 
ways in which a closed contour of length $L$ can be constructed by using $L$ edges.

We have $2N$ possible choices on where to put the first edge; we will call 
this choice step $1$. Since we have to build a closed curve, each of the 
two ending of the first edge has to be connected to other edges and each 
new edge can be connected in $3$ different ways to the previous one. We now 
proceed by iteration: at step $n\ge 2$ we add two new edges to the two 
open endings of the curve obtained at step $n-1$. Once the edge at step $1$ 
is fixed we have at most $3^{2(n-1)}$ way to build up the curve up to step $n$.
The length of the curve at step $n$ is 
$2(n-1)+1$ and when we arrive to the step $\bar{n}$ defined by 
\begin{equation*} 
2(\bar{n}-1)+1=L-1
\end{equation*}
(remember that $L$ is even) we have only one possible way to add the last edge 
to close the curve, so the number of closed curves of length $L$ 
has to be smaller than $2N3^{2(\bar{n}-1)}=2N3^{L-2}$. We now note 
that the same path can be obtained in a similar way by starting from a 
different edge in the first step, since all the edges of a closed curve 
are on the same footing, so we arrive to the upper bound
\begin{equation}\label{2d_len}
\#(L)\le \mathscr{N}(L)\equiv \frac{2N}{9L}3^L
\end{equation}
By using this estimate in \Eqref{2d_2} and remembering the 
definitions in \Eqref{2d_AL} and 
\Eqref{2d_X} we finally get
\begin{equation}\label{2d_3}
\langle N_-\rangle \le \sum_{L\ge 4, \mathrm{even}} A(L) \mathscr{N}(L) X(L)\le \frac{N}{72} \sum_{L\ge 4, \mathrm{even}} L\, 3^Le^{-2J\beta L}\ .
\end{equation}
This sum is convergent provided $3e^{-2J\beta}<1$ and the sum can be performed analytically 
(see appendix \ref{2dsum}). The final result is ($x$ is defined in \Eqref{x})
\begin{equation}\label{2d_final}
\langle N_-\rangle \le \frac{N}{36} x^2 \frac{2-x}{(1-x)^2}\ ,
\end{equation}
which is a bound of the form \Eqref{generalbound}.

\section{The $D=3$ problem}

We now consider a three dimensional square lattice of size 
$N^{1/3}\times N^{1/3}\times N^{1/3}$ 
with lattice spacing $a=1$. The Peierls contours of the two dimensional case 
now become surfaces, but their construction proceed along the same 
line as in the two dimensional case:
\begin{enumerate}
\item draw a unit cube on each site $i$ such that $s_i=-1$
\item cancel the faces that appear twice (\ie that separate two neighbour sites $i,j$ with $s_i=s_j=-1$)
\item if ambiguities are present, chop off the corners of the cubes in order to remove them.
\end{enumerate}
When using the $s_{i_b}=+1$ boundary condition we have the following properties
\begin{itemize}
\item every Peierls surface is a closed non-intersecting surface
\item every site $i$ with $s_i=-1$ is inside at least one surface
\item the set of the admissible surfaces is in a one-to-one
      correspondence with the set of the configurations.
\end{itemize}
which are the natural extension of the properties seen in the two dimensional case.

The bound in \Eqref{2d_0} becomes now 
\begin{equation}\label{3d_0}
N_-\le \sum_{S\ge 6, \mathrm{even}} \sum_{i=1}^{\#(S)} A(\gamma_{S}^i)X(\gamma_S^i)\ ,
\end{equation}
where $\gamma_S^i$ denotes the general Peierls surface composed by $S$ elementary squares, 
$A(\gamma_S^i)$ is the volume of $\gamma_S^i$ (that is the number of sites it 
contains) and $\#(S)$ is the number of closed surfaces of area $S$. $X(\gamma_S^i)$ is defined as 
in the two dimensional case, the lower extremum of the sum on $S$ is $6$ since this is the 
smallest area of a closed surface in $D=3$ and the sum extends on even numbers only since the area 
of a close surface is always even.

To show that, for sufficiently low temperature, a spontaneous magnetization is present 
we have to found the three dimensional analogs of the bounds \Eqref{2d_AL}, \Eqref{2d_X} and 
\Eqref{2d_len}.

\begin{figure}[t]
\begin{center}
\includegraphics[width=0.4 \textwidth, clip]{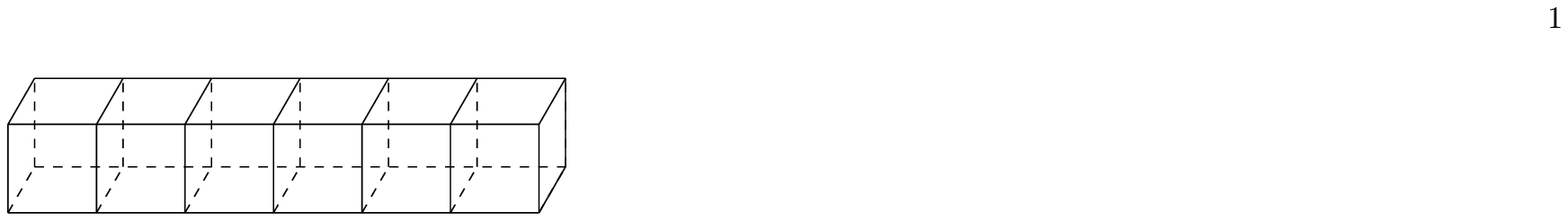}
\end{center}
\caption{A surface with the maximum $x_1$ value at fixed $S$: $x_1=(S-2)/4$.} \label{3d_AL_max}
\end{figure}

To find the three dimensional version of \Eqref{2d_AL} the procedure used in the two dimensional case has to be
slightly modified. Let us consider the smallest rectangular parallelogram $\mathcal{R}$ 
that contains the surface $\gamma_S^i$. In the three dimensional space it is not true that the area of $\mathcal{R}$ 
is not larger than the area of $\gamma_S^i$ (a simple counterexample is a donut shaped surface of 
sufficiently large radius), so we need a different constraint to be imposed on the edges $x_1, x_2, x_3$ of $\mathcal{R}$.
We can for example notice that we must have $x_1, x_2, x_2\le S/4$: by using $S$ elementary squares
to construct a closed surface, the maximum value we can get for $x_1$ (or $x_2$ or $x_3$) 
is $(S-2)/4$, which correspond to the surface shown in \Figref{3d_AL_max}. Thus we get
\begin{equation}\label{3d_AL}
A(\gamma_S^i)\le \max (x_1x_2x_3)\le (\max x_1)(\max x_2)(\max x_3) = A(S)\equiv \frac{S^3}{4^3}
\end{equation}
Proceeding in this way in the two dimensional case we would have obtained $A(L)=L^2/4$, which is
weaker than \Eqref{2d_AL}. 

The bound in \Eqref{2d_X} becomes 
\begin{equation}\label{3d_X}
\langle X(\gamma_S^i)\rangle \le X(S)\equiv e^{-2J\beta S}
\end{equation}
and the proof is completely analogous to one given for the two dimensional case: the transformation
$c\to\bar{c}$ now flip all the variables associated to the sites inside the surface $\gamma_S^i$ and 
\Eqref{2d_energy} becomes
\begin{equation}\label{3d_energy}
E(c)=E(\bar{c})+2JS\ .
\end{equation}

The last ingredient we need is the bound on $\#(S)$. Again we can proceed analogously to the
two dimensional case, by enumerating the possible ways in which we can put together $S$
elementary squares to obtain a closed surface. In the step number one we have 
$3N$ possible choices, and the surface is then built in the following way: at step $n$ we 
add $s_n$ squares to the surface, in such a way to saturate all the free edges present in the step $n-1$.

Here a little complication arises: in the planar case at every step we always  
have to add two more elements to the construction, while in the three dimensional case the number of elementary squares 
to be added is not constant in $n$, and in fact for some configuration this number is not even uniquely 
determined (\ie it depends on the way the squares are added). This complication is however not serious: 
an elementary square can be connected to a free edge in at most $3$ different ways, 
so at step $n$, when $s_n$ squares are added, we have at most $3^{s_n}$ possibilities. As a 
consequence, if the construction of the surface is completed in $\bar{n}$ steps, the total 
number of different possibilities is at most $3N 3^m$, where 
$m\equiv\sum_{n=2}^{\bar{n}}s_n=S-1$, independent of the construction details. As in
the two dimensional case, in this reasoning we overestimated the total number of 
different configurations by a factor $S$, since all the squares of a 
surface can be used as a starting point its construction. The final bound is thus:
\begin{equation}\label{3d_len}
\#(S)\le \mathscr{N}(S)\equiv N \frac{3^S}{S}\ .
\end{equation}

>From \Eqref{3d_0} we can now obtain an upper bound of the form in \Eqref{2d_3}:
\begin{equation}\label{3d_1}
\langle N_-\rangle \le \sum_{S\ge 6, \mathrm{even}} A(S)\mathscr{N}(S)X(S)=\frac{N}{64}\sum_{S\ge 6, \mathrm{even}}^{\infty} S^2\left(3e^{-2J\beta}\right)^S 
\end{equation}
This series is convergent provided $x=9e^{-4J\beta}<1$ 
and the sum can be performed analytically, see 
appendix \ref{3dsum}, the final result of the computation being
\begin{equation}\label{3d_final}
\langle N_-\rangle \le \frac{N}{16} x^3\frac{9-11x+4x^2}{(1-x)^3}\ .
\end{equation}

\section{The $D\ge 3$ problem}

The general case of $D$ dimensions, with $D>3$, does not present additional difficulties
with respect to the three dimensional setting studied before. We just need to substitute
``cube'' with ``hypercube'', ``face'' with ``hyperface'' and ``edge'' with ``hyperedge''.

The hyperface and hyperedge terms are not conventional but we will use them in order to make 
evident the similarity with the three dimensional case. An hyperface is an elementary $D-1$ 
surface in the $D$ dimensional space, 
\ie an hypercube in $D-1$ dimensions. In an analogous way an hyperedge is an 
elementary $D-2$ surface in the $D$ dimensional environment, \ie an hypercube in the $D-2$ 
dimensional space.

We will consider an hypercubic lattice of linear size $N^{1/D}$ and lattice spacing $a=1$. 
The construction of the Peierls domains proceeds along the same way as in  
$D=3$ and the Peierls domanis will now be closed non-intersecting hypersurfaces. 
Again each site $i$ with $s_i=-1$ is inside at least a Peierls hypersurface and, when the $s_{i_b}=+1$
conditions are imposed on the lattice boundary, the set of the admissible hypersurfaces is in a 
one-to-one correspondence with the set of the configurations.

The generalization of \Eqref{3d_0} is 
\begin{equation}\label{Dd_0}
N_-\le \sum_{H\ge 2D,\mathrm{even}} \sum_{i=1}^{\#(H)} A(\gamma_{H}^i)X(\gamma_H^i)\ ,
\end{equation}
where $H$ is the hyperarea of the Peierls surface $\gamma_H^i$. Again $H$ has to be even
and the smallest possible value for $H$ is $2D$.

To estimate $A(\gamma_H^i)$, as in the $D=3$ setting, we have to find the ``more elongated''
closed hypersurface composed of $H$ hyperfaces. This is given by the $D$ dimensional generalization
of \Figref{3d_AL_max}, for which:
\begin{equation*}
x_1\le \frac{H-2}{2(D-1)}\ .
\end{equation*}
As a consequence the bound in \Eqref{3d_AL} becomes 
\begin{equation}\label{Dd_AL}
A(\gamma_H^i)\le A(H)\equiv \left(\frac{H}{2(D-1)}\right)^D\ .
\end{equation}

The proof of the bound \Eqref{3d_X} goes on without significant modifications and the final result 
is again
\begin{equation}\label{Dd_X}
\langle X(\gamma_H^i)\rangle \le X(H)\equiv e^{-2J\beta H}\ .
\end{equation}

To get an estimate of $\#(H)$ we just have to notice that, as in the $D=3$ case, in order to build an 
hypersurface, an hyperface can be connected to a given hyperedge in no more than $3$ ways. As a consequence
also this estimate goes along the same lines as the one in $D=3$, the only difference being that in the 
first step we now have $DN$ possibilities instead of $3N$, obtaining
\begin{equation}\label{Dd_len}
\#(H) \le \mathscr{N}(H)\equiv DN\frac{3^H}{3H}\ .
\end{equation}

The final bound on $\langle N_-\rangle$ for the $D$ dimensional problem is thus
\begin{equation}\label{Dd_1}
\eqalign{
 \langle N_-\rangle & \le \sum_{H\ge 2D, \mathrm{even}} A(H)\mathscr{N}(H)X(H) \le \\
& \le \sum_{H\ge 2D, \mathrm{even}} \left(\frac{H}{2(D-1)}\right)^D \frac{DN}{3H}3^H e^{-2J\beta H}\ ,
}
\end{equation}
which, using the substitution $H=2k$ and \Eqref{x}, becomes
\begin{equation*}
\langle N_-\rangle \le \frac{ND}{6(D-1)^D}\sum_{k=D}^{\infty}k^{D-1}x^k \ .
\end{equation*}
For general $D$ the sum cannot be performed in a closed rational form, but, with the substitution 
$k=p+d$, can be rewritten as
\begin{equation*}
\sum_{k=D}^{\infty} k^{D-1}x^k=x^D\sum_{p=0}^{\infty}\frac{x^p}{(p+D)^{1-D}}=x^D\Phi(x,1-D,D)\ ,
\end{equation*}
where $\Phi$ is the Lerch transcendent function (see \eg \cite{BatemanBook} \S 1.11, \cite{DLMF}). 
The final result is thus
\begin{equation}\label{final}
\langle N_-\rangle  \le \frac{ND}{6(D-1)^D}x^D\Phi(x,1-D,D)\ ,
\end{equation}
which is of the form \Eqref{generalbound} since the Lerch transcendent is regular for $x=0$.

\begin{figure}[t]
\begin{center}
\includegraphics[width=0.45 \textwidth, clip]{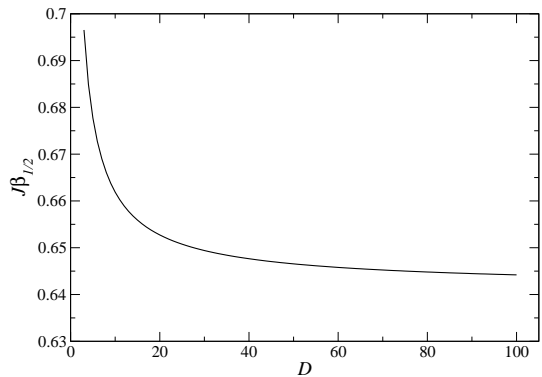}
\end{center}
\caption{Plot of $J\beta_{1/2}$ for $3\le D\le 100$.}\label{Jbeta_fig}
\end{figure}

\section{Conclusions}

We presented an elementary discussion of the Peierls argument
for the case of the $D-$dimensional ($D\ge 2$) Ising model defined on a cubical lattice.
The outcome of this arguments is an upper bound for $\langle N_-\rangle/N$
(Eqs.~(\ref{2d_final}), (\ref{3d_final}) and (\ref{final})), which implies for low enough temperature
a non-vanishing lower bound for the spontaneous magnetization and thus the presence of spontaneous 
symmetry breaking.

By looking back at the previous exposition we see that, apart from the numerical details, what makes the argument
sound is the fact that the ``entropy'' terms $A(H)$ and $\mathscr{N}(H)$ grow 
as a finite power of $H$, while the ``energy'' term $X(H)$ is exponentially dumped by $H$.
As a consequence the series in \Eqref{Dd_1} is convergent and the sum vanishes in the large $\beta$ limit.

The failure of this condition is the reason why the 
argument cannot be applied to the one dimensional Ising model: in that case the domains are just segments, 
$A(H)$ and $\mathscr{N}(H)$ still grow with $H$ but now $X(H)$ is $H-$independent (it is just $e^{-4\beta J}$). 
The upper bound for $\langle N_-\rangle$ is now a series which is divergent in the thermodynamical limit
and thus useless. In fact the one dimensional Ising model
can be analytically solved and no symmetry breaking is found for any positive value of the 
temperature (see \eg \cite{HuangBook}).

As a last remark we note that from the bound in \Eqref{final} we can get a bound for the critical coupling $\beta_c$, 
which is defined as the coupling at which the system switches from a ferromagnetic state to a 
paramagnetic state. From \Eqref{mmean} we see that, as far as $\langle N_-\rangle/N< 1/2-\epsilon$,
the system has to be ferromagnetic, so the critical value $x_c\equiv x(\beta_c)$ must lie outside the 
region $[0,x_{1/2}]$, where $x_{1/2}$ is defined as the smallest positive solution of the equation
\begin{equation*}
\frac{D}{6(D-1)^D}x^D\Phi(x,1-D,D)= \frac{1}{2}\ .
\end{equation*}
From $x_c>x_{1/2}$ we get $J\beta_c\le J\beta_{1/2}$, where 
\begin{equation*}
J\beta_{1/2}=\frac{1}{4}\log\left(\frac{9}{x_{1/2}}\right)
\end{equation*}
and some numerical values for $J\beta_{1/2}$ as a function of $D$ are reported in \Figref{Jbeta_fig}. For
comparison, the best available determination of the critical point for the three dimensional
Ising model is $J\beta_c(3D)=0.22165452(8)$ (see Ref.~\cite{DengBlothe}), from which we see that, for $D=3$, $J\beta_{1/2}$
is of the same order of magnitude of $J\beta_c$. For larger $D$ values this is however no more true.

The behaviour of the critical temperature of the Ising model in the limit of large $D$ is quite well known:  both lower 
\cite{Griffiths67, Fisher67} and upper \cite{FSS} bounds are known for $\beta_c$ and, by the combination of these bounds, 
one gets (\cite{DLP})
\begin{equation}\label{larged}
J\beta_c=\frac{1}{2D-1+\mathcal{O}(1/D)}\ ,
\end{equation}
and thus $\beta_c$ goes to zero as $D$ grows. On the other hand 
from \Figref{Jbeta_fig} we see that $\beta_{1/2}$ converges to a non-vanishing 
limit as $D\to\infty$. This is a well known limitation of the Peierls argument
(see the introduction of \cite{LebowitzMazel} for a discussion), 
which has to be modified in a non-elementary way to obtain an upper bound for $J\beta_{c}$ that 
vanishes in the large $D$ limit (\cite{LebowitzMazel}).

\section{Aknowledgments}

It is a pleasure to thank Pasquale Calabrese, Maurizio Fagotti and Silvia Musolino for
useful comments and discussions.

\appendix

\section{The sum in \Eqref{2d_3}}\label{2dsum}

By changing variable from $L$ to $n=L/2$ we get
\begin{equation*}
\sum_{L=4, \mathrm{even}}^{\infty} L\, 3^Le^{-2J\beta L}
=2\sum_{n=2}^{\infty} n(9e^{-4 J\beta })^n\ .
\end{equation*}
We introduce the variable $x$ defined in \Eqref{x} and, by using
$\sum_{n=0}^{\infty}x^n=1/(1-x)$, we obtain
\begin{equation*}
\eqalign{
2\sum_{n=2}^{\infty} n x^n &=2\left(\sum_{n=1}^{\infty}nx^n\right)-2x
=2x\frac{\rmd}{\rmd x}\left(\sum_{n=0}^{\infty}x^n\right)-2x=\\
&=2x\left(\frac{1}{(1-x)^2}-1\right)=2x^2\frac{2-x}{(1-x)^2}\ ,
}
\end{equation*}
which is the result used in the text.

\section{The sum in \Eqref{3d_1}}\label{3dsum}

By changing variable from $S$ to $k=S/2$ we have
\begin{equation*}
\sum_{S=6, \mathrm{even}}^{\infty}S^2\left(3 e^{-2J\beta}\right)^S=
4\sum_{k=3}^{\infty}k^2\left(9 e^{-4J\beta}\right)^k
=4\left(\sum_{k=1}^{\infty}k^2 x^k\right) -4x -16x^2\ ,
\end{equation*}
where in the second line we used the definition \Eqref{x}. Moreover
\begin{equation*}
\sum_{k=1}^{\infty}k^2 x^k=x\frac{\rmd}{\rmd x}\sum_{k=1}^{\infty}kx^k=
\left(x\frac{\rmd}{\rmd x}\right)^2\sum_{k=0}^{\infty}x^k\ ,
\end{equation*}
and by using $\sum_{n=0}^{\infty}x^n=1/(1-x)$ we obtain
\begin{equation*}
\sum_{k=1}^{\infty}k^2 x^k=\frac{x(1+x)}{(1-x)^3}\ .
\end{equation*}
The final result is thus
\begin{equation*}
\sum_{S=6, \mathrm{even}}^{\infty} S^2\left(3 e^{-2J\beta}\right)^S=
4x^3\,\frac{9-11x+4x^2}{(1-x)^3}\ .
\end{equation*}

\end{document}